\documentclass[prb,twocolumn,showpacs,superscriptaddress,amsmath,amssymb,reprint]{revtex4-1}

\usepackage{graphicx}
\usepackage{latexsym}
\usepackage{amsmath}
\usepackage{amssymb}
\usepackage{amsfonts}
\usepackage{color}
\usepackage[english]{babel}
\usepackage{bm}

\begin{document}
\bibliographystyle{apsrev}

\title{Fermi arcs formation in Weyl semimetals: the key role of intervalley interaction}

\author{Zh.A. Devizorova}
\affiliation{Moscow Institute of Physics and Technology, 141700 Dolgoprudny, Russia}
\affiliation{Kotelnikov Institute of Radio-engineering and Electronics RAS, 125009 Moscow, Russia}
\author{V.A. Volkov}
\affiliation{Kotelnikov Institute of Radio-engineering and Electronics RAS, 125009 Moscow, Russia}
\affiliation{Moscow Institute of Physics and Technology, 141700 Dolgoprudny, Russia}

\date{\today}
\begin{abstract}
We propose an analytical model describing Fermi arc surface states observed in the recent investigations of Weyl semimetals. The effective two-valley Hamiltonian is supplemented by the boundary conditions taking into account both the intravalley and intervalley interfacial interaction. We demonstrate that the latter is crucial for the formation of the surface states having the form consistent with the experimental data. Depending on the magnitude and interplay between the intravalley and intervalley interactions, the Fermi arc connects two nearby or distant valleys. Moreover, the emergence of additional Fermi contours (closed curves not intersecting the Weyl points) can be understood in the simplest four-valley approximation. These results open the opportunities for search of new effects in Weyl semimetals under an external fields.
\end{abstract}


\maketitle

\section{Introduction}
Weyl semimetals have attracted considerable attention recently as they extend the topological classification of matter beyond the insulators and demonstrate exotic surface states. The conduction and valence electron bands in these materials touch each other at special points of the Brillouin zone (BZ), called Weyl points, and the spectrum of quasiparticles is linear near these points. The stability of a Weyl point, which is the topological object due to a monopole-like structure of the Berry curvature, can be quantified by the corresponding chiral charge of this point \cite{Wehling}. 

Weyl semimetals demonstrate unusual surface states, the so-called Fermi arcs, whose Fermi contours take the form of arcs connecting two projections of the bulk Weyl points with the opposite chiral charges in the surface BZ \cite{Wan_PRB}. The non-trivial topology of the Berry curvature also guarantees the existence of Fermi arcs. Fermi arc states can lead to a new type of quantum oscillations due to a peculiar path of electrons in real and momentum space under an external magnetic field \cite{Potter_NatComm}, spin polarization textures \cite{Ojanen_PRB}, and unusual quantum interference effects in tunneling spectroscopies \cite{Hosur_PRB}. During the past few years, Weyl semimetals were intensively studied theoretically \cite{Murakami_NJP, Wan_PRB, Young_PRL, Yang_PRB, Xu_PRL, Burkov_PRL, Burkov_PRB, Halasz_PRB, Huang_NatComm}, and recently both the surface Fermi arcs and the bulk cones have been experimentally observed in the Weyl semimetals TaAs and NbAs \cite{Xu_Science, Lv_NatPhys, Yang_NatPhys,Xu_NatPhys}.

Even in the single-valley approximation the surface states in Weyl semimetals can be quite peculiar: band bending near the crystal boundary gives rise to a spiral structure of its Fermi contours \cite{Li_PRB}. Li and Andreev in Ref.[\onlinecite{Li_PRB}] suggested that the Fermi arc may be understood in terms of the avoided crossing of two spirals belonging to the close valleys. However, according to the experimental data \cite{Xu_Science, Lv_NatPhys, Yang_NatPhys, Xu_NatPhys}, the filling of both bulk and the surface states terminates near the neutrality point, thus significant band bending is hardly present. Despite the first-principles calculations show the existence of the Fermi arc surface states in TaAs and NbAs \cite{Huang_NatComm, Xu_NatPhys, Sun_PRB} and these states were experimentally observed \cite{Xu_Science, Xu_NatPhys, Lv_NatPhys, Yang_NatPhys}, there still exists no simple analytical model of the Fermi arcs not relying on additional mechanisms and taking into account only the non-trivial topology of Weyl semimetals and sharp discontinuity of the crystal potential at the sample surface. 

\begin{figure}[b!]
\center{\includegraphics[width=0.6\linewidth]{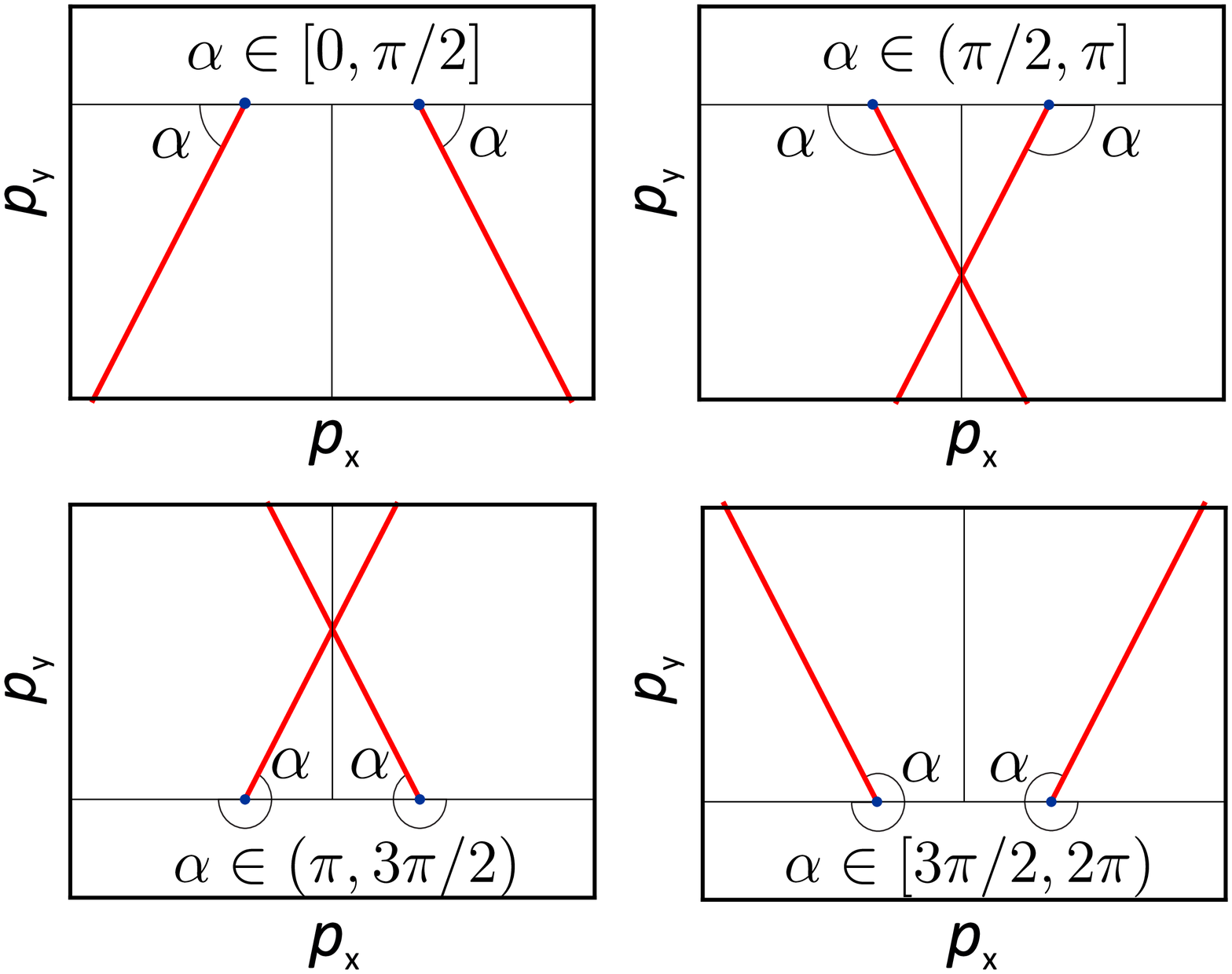}}
\caption{(Color online) The sketches of the (001) surface states' Fermi contours at $E=0$ in the absence of the intervalley interaction ($\gamma=0$) for different intravalley boundary parameters $\alpha$. The blue dots are the projections of the bulk Weyl points.} \label{Fig_1}
\end{figure}

\begin{figure}[t!]
\center{\includegraphics[width=0.6\linewidth]{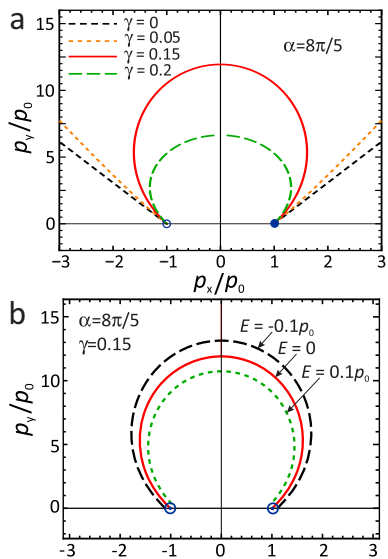}}
\caption{(Color online) (a) The evolution of the (001) surface states' Fermi contours as a function of the intervalley interaction $\gamma$ for the fixed intravalley parameter $\alpha$ corresponding to not intersecting rays (see Fig. \ref{Fig_1}) in the limit $\gamma=0$. The empty and solid blue circles show the projections of the bulk Weyl points with the opposite chiral charges. (b) The modification of the Fermi arc with varying energy. Here the blue circles are the projections of the bulk Weyl cones and Weyl points.} \label{Fig_2}
\end{figure}

\begin{figure}[t!]
\center{\includegraphics[width=0.6\linewidth]{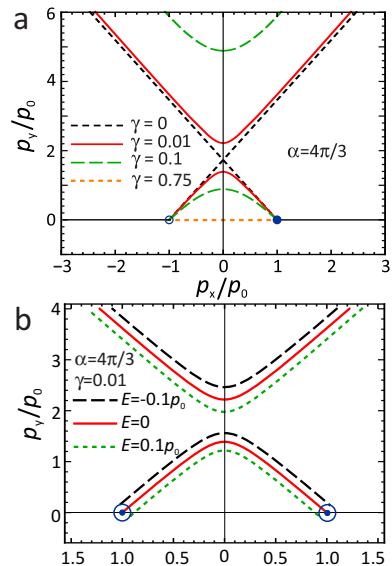}}
\caption{(Color online) (a) The (001) surface states Fermi contours depending on the intervalley interaction $\gamma$ at fixed $\alpha$, for which the Fermi rays have the crossing point in the limit $\gamma=0$. (b) The evolution of the (001) surface states' Fermi surfaces with varying energy $E$.}\label{Fig_3}
\end{figure}

In this paper, we propose such a model. The bulk spectrum with two Weyl points is described by the effective two-valley $kp$ Hamiltonian. To describe the surface states, the Hamiltonian must be considered together with the boundary conditions on the sample surface. We derive the boundary conditions from general physical requirements. Previously, such approach was successfully applied to the description of the surface states in the materials with Dirac spectrum \cite{Volkov}, graphene \cite{McCann_JPCM, Akhmerov_PRB, Zagorodnev_FNT, Basko_PRB, Tkachov_PRB, Ostaay_PRB} and topological insulators \cite{Enaldiev_PZhETF, Menshov_JETPL}. Both the intravalley and intervalley interaction arising due to the atomically sharp interfacial potential are taken into consideration. Next, we analyze the surface states spectra. In the case of noninteracting valleys, the Fermi contours take the form of rays which can either intersect or not (see Fig. {\ref{Fig_1}}). In the latter case, the presence of the intervalley interaction above some threshold value leads to the linking of the rays with the Fermi arcs formation, as shown in  Fig. \ref{Fig_2}a. In the former case, the rays repel at the crossing point due to the intervalley interaction also forming the Fermi arcs (see Fig. \ref{Fig_3}a). The Fermi contours which do not form arcs in the two-valley approximation can nevertheless connect two remote valleys. We consider this possibility qualitatively in the four-valley approximation. 

We start with the two-valley approximation in which the electron wave function with the energy $E$ obeys the Dirac-like equation
\begin{equation}
\label{WE}
\left( \begin{array}{cc}
        {\bm \sigma}(\hat {\bf p}+{\bf p}_0) & \Delta \sigma_x\\
      \Delta \sigma_x &  - {\bm \sigma}(\hat {\bf p}-{\bf p}_0)
    \end{array} \right) \left( \begin{array}{cc}
      \psi \\
      \phi
    \end{array} \right)=E \left( \begin{array}{cc}
      \psi \\
      \phi
    \end{array} \right),
\end{equation}
where  $\psi=(\psi_1,\psi_2)^T$, $\phi=(\phi_1,\phi_2)^T$ are the two-component pseudospinors, ${\bf p}=(\hat p_x,\hat p_y,\hat p_z)$ is the momentum operator, $ {\bm \sigma}=(\sigma_x,\sigma_y, \sigma_z)$ is the set of Pauli matrices, $\Delta$ characterizes the intervalley interaction in the bulk, ${\bf p}_0=(p_0,0,0)$ determines the position of the Weyl points without the bulk intervalley interaction. We assume the Weyl velocity to be $v=1$.

We consider the semi-infinite ($z \ge 0$) system and introduce the boundary condition in the following form
\begin{equation}
\label{BC_gf}
(\psi+i\hat g \phi)|_{z=0}=0,
\end{equation}
where $\hat g$ is unknown matrix. Our next goal is to find the matrix $\hat g$. The first constraint on it is impossed by the requirement of the Hamiltonian Hermiticity in the bounded region. Using this requirement we obtain the following boundary condition for arbitrary effective wave functions 
\begin{equation}
(\psi^{\dagger}_{\lambda} {\bm \sigma} {\bf n} \psi_{\nu} -\phi^{\dagger}_{\lambda}{\bm \sigma} {\bf n} \phi_{\nu})|_{z=0}=0,
\end{equation}
where ${\bf n}$ is the unit vector normal to the boundary $z=0$. Combining it with ({\ref{BC_gf}}), we obtain $g^+{\bm \sigma} {\bf n} g={\bm \sigma} {\bf n}$.

To further reduce the number of the parameters in $g$, we use additional symmetry arguments. If only two closed valleys connected by the Fermi arc are considered, then the experimentally observed spectra of both the bulk and surface states are symmetric with respect to the inversion $p_x \rightarrow -p_x$ or $p_y \rightarrow -p_y$, depending on the location of the valleys \cite{Xu_Science, Xu_NatPhys}. For the chosen position of the valleys, we assume the symmetry $x \rightarrow -x$ to be present in the system. Due to the fact that the operator $\hat I$ corresponding to this symmetry commutes with the Hamiltonian, i.e.  $[\hat H, \hat I]=0$, it takes the form
\begin{equation}
\hat I=\left( \begin{array}{cc}
        0 &  \sigma_x\\
       \sigma_x &  0
    \end{array} \right) \hat i_{x \rightarrow -x},
\end{equation}
where $\hat i_{x \rightarrow -x}$ is the operator of the replacement of $x$ to $-x$. The boundary condition ({\ref{BC_gf}}) must be symmetric with respect to the transformation $\hat I$, thus $\sigma_x g^{-1}\sigma_x=-g$.

Combining the two constraints for the matrix $g$, we write it down in the form
\begin{equation}
\label{g}
g(\alpha, \gamma)=\frac{i}{\sqrt{\gamma}}\left( \begin{array}{cc}
      e^{-i\alpha} & \mp i\sqrt{1-\gamma}\\
       \pm i\sqrt{1-\gamma} & e^{i\alpha}
    \end{array} \right),
\end{equation}
where $\alpha \in [0,2\pi)$ and $\gamma \in (0,1)$ are two phenomenological parameters depending on the bulk band structure as well as on the microscopic structure of the interface. Our boundary condition differs from that derived in Ref.[\onlinecite{Li_PRB}]. The values of the boundary parameters can be extracted from comparison with experiments, as it was done for the edge states in graphene \cite{Latyshev_SciRep}. The transition from the boundary condition where $g$ is taken with upper sign to the boundary condition with the lower sign is realized by shifting $\alpha$ by $\pi$. Further, for certainty we consider only the upper sign in $g$, because $\alpha$ is a free parameter. 

In the limiting case $\gamma \rightarrow 0$, the wave functions in the different valleys are decoupled, and we obtain the single-valley boundary conditions in the following form

\begin{gather}
\label{BC_SV}
(\phi_1+e^{i(\alpha-\pi/2)}\phi_2)|_{z=0}=0, \\
(\psi_1+e^{-i(\alpha-\pi/2)}\psi_2)|_{z=0}=0.
\end{gather}

We conclude that the parameter $\alpha$ determines the surface states spectrum in the isolated valleys, while $\gamma$ characterizes the interfacial intervalley interaction. The boundary condition similar to (\ref{BC_SV}) was already derived in the single valley approximation \cite{Li_PRB, Hashimoto_arxiv}. However, in such approach the intravalley boundary parameter in one valley is generally different from the parameter in the other. We show that these parameters are not independent.

Now we obtain the (001) surface states spectra. There are two mechanisms of the intervalley interaction: bulk (described by $\Delta$) and interfacial (characterized by $\gamma$). The bulk valley coupling itself does not lead to the formation of a Fermi arc. Aiming to analyze the influence of the abrupt (in the atomic scale) interfacial potential we neglect $\Delta$. Substituting the solution of the equation (\ref{WE}) with $\Delta=0$ into the boundary condition (\ref{BC_gf}) with the matrix $g$ satisfying (\ref{g}), we obtain the system of dispersion equations

\begin{multline}
\label{DE1}
\sqrt{1-\gamma}[E(\hbar \kappa_--\hbar \kappa_+)+2p_xp_y]-\\-\cos \alpha [2Ep_x+(\hbar \kappa_- -\hbar \kappa_+)p_y]+\\+\sin \alpha [\hbar \kappa_+ (p_x-p_0)+\hbar \kappa_- (p_x+p_0)]=0,
\end{multline}

\begin{multline}
\label{DE2}
\sqrt{1-\gamma}(E^2+\hbar^2 \kappa_- \kappa_+-p_x^2+p_0^2+p_y^2)-\\- \cos \alpha [2Ep_y+\hbar \kappa_+ (p_x-p_0)-\hbar \kappa_-(p_x+p_0)]+\\+ \sin \alpha [2Ep_0+(\hbar \kappa_+ +\hbar \kappa_-)p_y]=0,
\end{multline}
where $\hbar \kappa_{\pm} =\sqrt{({\bf p} \pm {\bf p_0})^2-E^2}$.

It is more illustrative to analyze not the spectrum, but the surface states' Fermi contours at different energies. First, we consider in detail the zero energy Fermi surfaces. We introduce the elliptic coordinates
\begin{equation}
\label{px_py}
p_x=p_0\cosh u \cos v, \qquad p_y=p_0\sinh u \sin v.
\end{equation}

The system (\ref{DE1})-(\ref{DE2}) is equivalent to one equation
\begin{equation}
\label{DE_EC}
\sin v(\sqrt{1-\gamma} \cosh u+\cos \alpha)= -\sin \alpha \sinh u,
\end{equation}
which couples the parameters $u$ and $v$. 

If $\gamma=0$, which corresponds to the absence of the intervalley interaction, the Fermi contours of the surface states are the rays emanating from the projections of the bulk Weyl points on the surface BZ. The relative orientation of the rays depends on the intravalley boundary parameter $\alpha$ (see Fig. \ref{Fig_1}).

If the intervalley interaction is taken into account, the surface states' Fermi contours can be introduced parametrically. In the case  $\sqrt{1-\gamma}\cosh u + \cos \alpha \ne 0$, the parameter $v$ is expressed via $u$ from (\ref{DE_EC}). 

If $\alpha \ne \pi$ and $\sqrt{1-\gamma} \ne -\cos \alpha$, all solutions are given by Eq. (\ref{px_py}), where $v$ is determined by (\ref{DE_EC}) and $u$ is a free parameter. In the case $\alpha=\pi$, apart from the solutions with $v$ from (\ref{DE_EC}), there exists an extra solution $p_x=p_0\cos v/\sqrt{1-\gamma}$, $p_y=p_0\sin v\sqrt{\gamma/(1-\gamma)}$, where $v \in [0,2\pi)$; if $\sqrt{1-\gamma}=-\cos \alpha$ the extra solution is $p_x=p_0\cos v$,  $p_y=0$,  $v \in [0,2\pi)$.

The Fermi surfaces are presented in Figs. \ref{Fig_2}a and \ref{Fig_3}a. In Fig. \ref{Fig_2}a the intravalley boundary parameter $\alpha$ corresponds to the case when in the absence of the intervalley interaction the rays do not intersect ($\cos \alpha \ge 0$). Next, the modification of the Fermi surfaces shape with increasing the intervalley interaction $\gamma$ is shown. The Fermi contours merge with each other forming the Fermi arc at $\gamma > \cos^2\alpha$. The Fermi arcs of such kind have been experimentally observed recently in several Weyl semimetals \cite{Xu_Science, Xu_NatPhys,Yang_NatPhys}. 

In Fig. \ref{Fig_3}a the rays have the crossing point at $p_x=0$ if $\gamma=0$. The introduction of an arbitrarily weak intervalley interaction leads to the anticrossing of the rays with the formation of two branches, one of which is the Fermi arc connecting two Weyl points. 

Comparing our Eqs.(\ref{px_py}--\ref{DE_EC}) with the experimental data presented in Fig.3b of Ref.[\onlinecite{Yang_NatPhys}], where two Fermi arcs connect two pairs of valleys with different $k_z$, we obtain the following values of the boundary parameters: $\alpha \approx 5.64$ radians, $\gamma \approx 0.88$ for the larger Fermi arc and $\alpha \approx 5$ radians, $\gamma \approx 0.44$ for the smaller one.

In the case of nonzero energy, the Fermi surfaces can be obtained from simultaneous solution of (\ref{DE1}) and (\ref{DE2}). The dependencies of the Fermi arcs shape on the energy for different boundary parameters are presented in Figs.\ref{Fig_2}b and \ref{Fig_3}b.

In the two-valley approximation, apart from the Fermi arc-like solutions, there exist some decoupled  Fermi surfaces. The solutions which do not form an arc connecting two close valleys also appear in the numerical simulations \cite{Huang_NatComm, Sun_PRB}. These uncoupled solutions emerge because we considered only two valleys located near the BZ center. In real materials the number of Weyl points is greater than two, and these points can be located close to the BZ edges \cite{Xu_Science, Xu_NatPhys, Lv_NatPhys, Yang_NatPhys, Sun_PRB}. For this reason, the constant-energy curve emanating at one of the Weyl points can terminate at the Weyl point located in another edge of BZ. To demonstrate such possibility qualitatively, we consider the model with two pairs of valleys located in the opposite edges of BZ
\begin{equation}
\hat H=\left( \begin{array}{cccc}
        {\bm \sigma}(\hat {\bf p}+{\bf p}_1) & 0 & 0 & 0\\
        0 &- {\bm \sigma}(\hat {\bf p}-{\bf p}_2) & 0 & 0 \\
        0 & 0& - {\bm \sigma}(\hat {\bf p}+{\bf p}_2) & 0\\
        0 & 0 & 0 &  {\bm \sigma}(\hat {\bf p}-{\bf p}_1)
    \end{array} \right),
\end{equation}
where ${\bf p}_1=(p_{0x},-p_{0y},0)$, ${\bf p}_2=(p_{0x}, p_{0y},0)$,  $p_{0x} \ll p_{0y}$.

The system is symmetric with respect to the inversions $x \rightarrow -x$ and $y \rightarrow -y$, the operators of the corresponding transformations are
\begin{gather}
\hat I_x=\left( \begin{array}{cc}
        I_1 & 0\\
       0 &  I_1
    \end{array} \right) \hat i_{x \rightarrow -x}, \qquad I_1=\left( \begin{array}{cc}
        0 & \sigma_x\\
       \sigma_x &  0
    \end{array} \right), \\
\hat I_y=\left( \begin{array}{cc}
        0 & I_2\\
       I_2 &  0
    \end{array} \right) \hat i_{y \rightarrow -y}, \qquad I_2=\left( \begin{array}{cc}
        \sigma_y & 0\\
       0 &  \sigma_y
    \end{array} \right).
\end{gather}

Exploiting the relation $p_{0x} \ll p_{0y}$, we assume that the boundary condition mixes the envelope functions in each pair of valleys, but does not mix functions of different pairs, i.e.
\begin{equation}
(\psi+i\hat g_1 \phi)|_{z=0}=0, \qquad (\psi '+i\hat g_2 \phi ')|_{z=0}=0,
\end{equation}
where $\psi$, $\phi$, $\psi '$, $\phi '$ are the two-component pseudospinors corresponding to the valleys, which Weyl points are located at $-{\bf p}_1$, ${\bf p}_2$, $-{\bf p}_2$, ${\bf p}_1$, respectively.

\begin{figure}[t!]
\center{\includegraphics[width=0.7\linewidth]{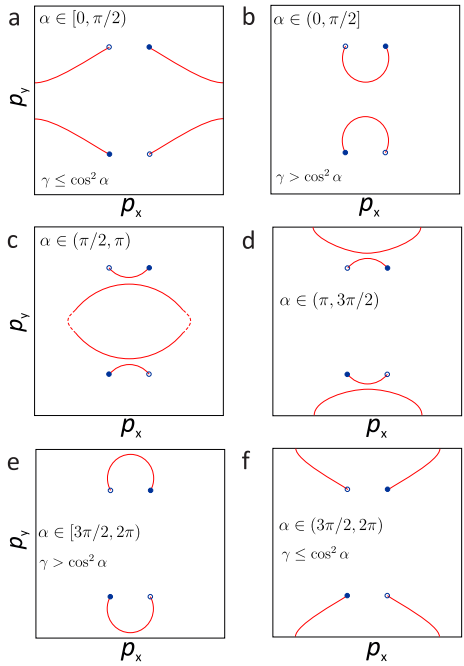}}
\caption{(Color online) The schematic view of the mutual orientation of the Fermi contours at $E=0$ in four-valley approximation.} 
\label{Fig_4}
\end{figure}

From the requirement of the Hermiticity in the half-space and symmetry with respect to the inversion $x \rightarrow -x$, it follows that $g_1=g(\alpha, \gamma)$, $g_2=g(\alpha ', \gamma ')$. Using the symmetry with respect to the inversion $y \rightarrow -y$ we obtain $\alpha '=-\alpha $, $\gamma '=\gamma$. The possible mutual orientations of the Fermi surfaces are shown in Fig.\ref{Fig_4}. Depending on the position of the Weyl points in BZ and the value of the boundary parameters, the rays which do not form a Fermi arc in the two-valley approximation can interact in the different ways. We note that the $kp$-approximation is valid just in the vicinity of Weyl points, and the demonstrated Fermi arcs in the whole BZ can be considered as schematic only. If $\alpha$ belongs to the second or third quadrant, then for each pair of valleys there exists an unclosed branch, which does not pass through the projections of the Weyl points. Such branches can also interact in different ways forming the closed curve not intersecting the Weyl points. These extra Fermi surfaces reflect how the Weyl points can annihilate during the transition from trivial to topological insulator (for details see Ref.[\onlinecite{Huang_NatComm}]).

Besides Weyl semimetals with pointlike Fermi surfaces (type I) there exist type II Weyl semimetals with strongly tilted Weyl cones \cite{Soluyanov_Nat}. It was shown recently that such semimetals demonstrate not only topological, but also trivial Fermi arcs \cite{Tamai_PRX}. Our results are not directly applicable to type II Weyl semimetals, because the latter are described by a more general Hamiltonian containing unit matrix in addition to Pauli matrices \cite{Soluyanov_Nat}. However, using the same methods one can derive the boundary conditions and surface states spectra for type II case. We anticipate that such model would describe both topological and trivial Fermi arcs, but this is the subject for further studies. 

In conclusion, we have developed an analytical model demonstrating the key role of the intervalley interactions in the formation of Fermi arc surface states in type I Weyl semimetals. We have obtained the general boundary condition for the effective wave functions on the surface of a Weyl semimetal in the two-valley approximation. It contains two real phenomenological parameters. One describes the intravalley and the other one describes the intervalley interfacial interaction. We show that the shape and connectivity of the surface states are determined by the interplay between these parameters. The interaction between two pairs of valleys has been qualitatively analyzed in the four-valley approximation. Our continuum model can be easily generalized for the presence of electric, magnetic, etc. fields, and drastically simplifies the analysis of their effect on the properties of Weyl semimetals.

This work is supported by the Russian Science Foundation (Project No. 16-12-10411).

\end{document}